\documentclass[aps,prev,preprintnumbers,floatset,nofootinbib,twocolumn]{revtex4-1}%twocolumn
\usepackage[english]{babel}
\usepackage{bm}
\usepackage{times}
\usepackage{ulem}
\usepackage{hyperref}
\hypersetup{
    colorlinks=true,
    linkcolor=blue,
    filecolor=magenta,      
    urlcolor=blue,
    citecolor=blue,
    pdftitle={Sharelatex Example},
    pdfpagemode=FullScreen
    }
\usepackage{listings}
\usepackage{slashed}
\usepackage{color}
\usepackage{appendix}
\usepackage{mathtools}
\usepackage{epsfig}
\usepackage{dcolumn}
\usepackage{textcomp}
\usepackage{appendix} 
\usepackage{multirow}
\usepackage{graphicx}
\usepackage{soul}
\usepackage{subfigure}
\newcommand{\fig}[1]{Fig.~\ref{#1}} 
\newcommand{\figs}[1]{Figs.~\ref{#1}} 
\newcommand{\eq}[1]{Eq.~(\ref{#1})} 
\newcommand{\tab}[1]{Table~\ref{#1}} 
\graphicspath{{PLots/}}

\newcommand{\Yoxara}[1]{{\color{magenta}#1}}

\begin{document} 

%\preprint{IIPDM-2021}

%\title{3-3-1 Models at High-Luminosity, High-Energy and the Future Circular Collider}
\title{Constraining 3-3-1 Models at the LHC and Future Hadron Colliders}

\author{A. Alves$^{1}$}
\email{aalves@unifesp.br}
\author{L. Duarte$^{7}$}
\email{l.duarte@unesp.br}
\author{S. Kovalenko$^{2,3,4,5}$}
\email{sergey.kovalenko@unab.cl}
\author{Y. M. Oviedo-Torres$^{6,7}$}
\email{ymot@estudantes.ufpb.br}
\author{F. S. Queiroz$^{3,7,8}$}
\email{farinaldo.queiroz@ufrn.br}
\author{Y. S. Villamizar$^{7,8}$}
\email{yoxara@ufrn.edu.br}

\affiliation{$^1$Departamento de F\'isica, Universidade Federal de S\~ao Paulo, UNIFESP, Diadema, S\~ao Paulo, Brazil}
\affiliation{$^2$Departamento de Ciencias Fisicas, Universidad Andres Bello,
Sazie 2212, Piso 7, Santiago, Chile}
\affiliation{$^3$ Millennium Institute for Subatomic Physics at High-Energy Frontier (SAPHIR), Fernandez Concha 700, Santiago, Chile}
\affiliation{$^4$ Centro Cient\'{\i}fico-Tecnol\'{o}gico de Valpara\'{\i}so, Casilla 110-V, Valpara\'{\i}so, Chile}
\affiliation{$^5$ Bogoliubov Laboratory of Theoretical Physics, JINR, 141980 Dubna, Russia}
\affiliation{$^6$ Departamento de Fisica, Universidade Federal da Paraiba, Caixa Postal 5008, 58051-970, Joao Pessoa, PB, Brazil}
\affiliation{$^7$ International Institute of Physics, Universidade Federal do Rio Grande do Norte, Campus Universitario, Lagoa Nova, Natal-RN 59078-970, Brazil}
\affiliation{$^8$ Departamento de F\'isica, Universidade Federal do Rio Grande do Norte, 59078-970, Natal, RN, Brasil}

\begin{abstract}
In this work, we derive lower mass bounds on the $Z^\prime$  gauge boson based on the dilepton data from LHC with 13 TeV of center-of-mass energy, and forecast the sensitivity of the High-Luminosity-LHC with $L=3000 fb^{-1}$, the High-Energy LHC with $\sqrt{s}=27$~TeV, and also at the Future Circular Collider with $\sqrt{s}=100$~TeV. We take into account the presence of exotic and invisible decays of the $Z^\prime$ gauge boson to find a more conservative and robust limit, different from previous studies. We investigate the impact of these new decay channels for several benchmark models in the scope of two different 3-3-1 models. We found that in the most constraining cases, LHC with $139fb^{-1}$ can impose $m_{Z^{\prime}}>4$~TeV. Moreover, we forecast HL-LHC, HE-LHC, and FCC-hh collider reach, and derived the projected bounds $m_{Z^{\prime}}>5.8$~ TeV, $m_{Z^{\prime}}>9.9$~TeV, and $m_{Z^{\prime}}> 27$~TeV, respectively. Lastly, we put our findings into perspective with dark matter searches to show the region of parameter space where a dark matter candidate with the right relic density is possible.
\noindent

\end{abstract}


\maketitle
\flushbottom

\section{\label{In} Introduction}

Neutral resonances decaying to lepton pairs occur in several beyond the Standard Model (SM) theories that are motivated to explain open problems such as dark matter (DM), neutrino masses, parity violation and grand-unification, for example. Many of such models predict neutral gauge bosons, which can be produced at current and future colliders. In some dark matter models \cite{Arcadi:2017kky}, $Z^\prime$ gauge bosons mediate interactions with the SM spectrum and are key to the dark matter phenomenology, as they can drive both the relic density and direct detection signals. A dark matter particle at the LHC is inferred from missing energy events, but the LHC can also indirectly contribute to the dark matter hunting by observing decays of those  gauge bosons that mediate the DM-SM interactions. The dilepton channel is particularly interesting since it is much cleaner than the di-jet one, offering a better signal-over-background ratio. \newline

From a collider physics perspective, obtaining mass bounds on a new vector boson is a promising strategy to assess which new physics models could be observed at the LHC. For instance, if a dilepton signal is observed with invariant mass at $4$~TeV, one could conclude from our findings that such signal would not come from a model based on the $SU(3)_C \otimes SU(3)_L \otimes U(1)_X$ gauge symmetry, 3-3-1 for short.  Models based on this symmetry are phenomenologicaly compelling because may solve some open problems such as neutrino masses \cite{Montero:2000rh, Tully:2000kk,Montero:2001ts,Cortez:2005cp,Cogollo:2009yi,Cogollo:2010jw,Cogollo:2008zc,Okada:2015bxa,Vien:2018otl,carcamoHernandez:2018iel,Nguyen:2018rlb,Pires:2018kaj,CarcamoHernandez:2019iwh,CarcamoHernandez:2019vih,CarcamoHernandez:2020pnh}, dark matter \cite{Fregolente:2002nx,Hoang:2003vj,deS.Pires:2007gi,Mizukoshi:2010ky,Ruiz-Alvarez:2012nvg,Profumo:2013sca,Dong:2013ioa,Dong:2013wca,Cogollo:2014jia,Dong:2014wsa,Dong:2014esa,Kelso:2014qka,Dong:2014esa,Mambrini:2015sia,Dong:2015rka,deSPires:2016kkg,Alves:2016fqe,RodriguesdaSilva:2014gbi,Carvajal:2017gjj,Dong:2017zxo,Arcadi:2017xbo,Montero:2017yvy,Huong:2019vej,Alvarez-Salazar:2019cxw,VanLoi:2020xcq,Dutra:2021lto,Oliveira:2021gcw}, meson anomalies \cite{Cogollo:2012ek,Cogollo:2013mga,Buras:2014yna,Buras:2015kwd,Queiroz:2016gif,deMelo:2021ers,Buras:2016dxz,Buras:2021rdg}, flavor violation \cite{Cabarcas:2012uf,Hue:2017lak}, among others \cite{Montero:2011tg,Santos:2017jbv,Barreto:2017xix,DeConto:2017ebj,Chen:2021haa,Chen:2021zwn}.

%Besides, they contain a new gauge boson  $Z^{\prime}$, which couple with SM particles that can leave a very clean signal in the detectors in specific channels like high boosted muons, a pair of leptons and missing transverse energy (${\mu}^{\pm}{\mu}^{\mp}{\ell}^{\mp}{\nu}_{\ell}{\ell}^{\pm}{\nu}_{\ell}$), a pair of leptons (like de dimuon channel ${\mu}^{\pm}{\mu}^{\mp}$) and 4 light jets \cite{cao2016collider}.

An important feature of these models is that the mass of the $Z^\prime$ gauge boson is determined by the energy scale at which the 3-3-1 symmetry breaks down to the 3-2-1.  Furthermore, masses of the particles belonging to the 3-3-1 spectrum are also proportional to this energy scale. In other words, a bound on the $Z^\prime$ mass is seen as a limit on the entire 3-3-1 spectrum. For concreteness, we will focus our analysis on two popular models based on the 3-3-1 symmetry, namely 3-3-1 RHN and 3-3-1 LHN. As the symmetry suggests, the fermion content is arranged in triplets under $SU(3)_L$. In the 3-3-1RHN, the lepton triplet contains a right-handed neutrino, whereas the 3-3-1 LHN has a heavy neutral fermion. The key difference between these models is the presence of a viable dark matter candidate. In the latter, such heavy neutral fermion can reproduce the correct dark matter relic density and yield signals at direct detection experiments via interactions mediated by the $Z^\prime$ field. Therefore, a limit on the $Z^\prime$ boson translates into a constraint on possible dark matter signals. Lastly, 3-3-1 models have $W^\prime$ bosons, whose mass is also set by the energy scale of symmetry breaking, i.e., by the $Z^\prime$ mass. These new gauge boson can induce lepton flavor violation signals \cite{Lindner:2016bgg,Arcadi:2017xbo}, which again can be indirectly constrained by lower mass bounds on the $Z^\prime$ mass. We point out that 3-3-1 models naturally induce FCNC (flavor changing neutral currents) processes because one of the fermion generations transform differently under $SU(3)_L$. The FCNC processes stems from the $Z^\prime$ and scalar fields. It has been shown that scalars yield relatively smaller FCNC interactions \cite{Cogollo:2013mga,Machado:2016jzb,Queiroz:2016gif}. That said, the $B_d$ meson system in particular, can lead to strong constraints on the $Z^\prime$ mass depending on the parametrization used for the mixing matrices. Nevertheless, colliders offer an orthogonal and cleaner probe. Anyway, the discussion of FCNC in the context of 3-3-1 models is out of our scope. Therefore, limit ourselves to collider physics.

Dedicated collider studies of $Z^\prime$ bosons in the context of 3-3-1 models have been carried out in the past. A projected limit of 600 GeV has been derived in a linear collider \cite{RamirezBarreto:2007cie}.
In \cite{RamirezBarreto:2010vji} the authors have derived the expected number of signal events for the LHC with $\sqrt{s}=7,14$~TeV. A similar study was done in \cite{Corcella:2017dns}, but focusing on a doubly charged vector boson present in some 3-3-1 models whose mass is connected to the $Z^\prime$. Hence, one could treat %see 
it as an indirect probe for $Z^\prime$ bosons.  In \cite{RamirezBarreto:2013edk} the authors derived an indirect lower mass bound on the $Z^\prime$ gauge boson using the mass relation between the $Z^\prime$ and the vector doubly charged gauge boson in a 3-3-1 model. In \cite{Coutinho:2013lta} two lower $Z^\prime$ mass bounds, $2.2$~TeV and $2.5$~TeV, were obtained from the LHC data using different 3-3-1 models. 
%In \cite{Coutinho:2013lta} two lower $Z^\prime$ mass bounds were obtained using LHC data %that read $2.2$~TeV and $2.5$~TeV again considering different 3-3-1 models. 
In \cite{Nepomuceno:2019eaz} a more updated analysis has been carried out but again focused on the doubly charged gauge boson. In \cite{Queiroz:2016gif}  and \cite{Lindner:2016bgg} the authors derived 
a lower mass limit, $3-4$~TeV, on the $Z^\prime$ boson considering only $Z^\prime$ decays into charged leptons.
%a lower mass limit on the $Z^\prime$ boson that reads $3-4$~TeV considering only %$Z^\prime$ decays into charged leptons. 
A discussion concerning the relevance of exotic $Z^\prime$ decays has been already raised in \cite{DeJesus:2020yqx}, but a solid calculation was still missing. Hence, it is clear that an updated and comprehensive derivation of lower mass limits on the $Z^\prime$ gauge boson belonging to 3-3-1 models was missing up to now. 

Motivated by the importance of the $Z^\prime$ gauge boson to 3-3-1 constructions, we compute lower mass bounds on $Z^\prime$ boson based on dilepton decays, $Z^\prime\to \ell^+\ell^-,\; \ell=e,\mu$, for both 3-3-1RHN and 3-3-1LHN models using 139 fb$^{-1}$ of data collected from proton-proton collisions at the 13 TeV~\cite{atlas2019search} taking into account overlooked exotic decays, such as decay into exotic quarks and dark matter. These new decay channels might significantly impact the lower bound obtained considering only decays into SM fermions as previously assumed \cite{RamirezBarreto:2007cie,RamirezBarreto:2010vji,RamirezBarreto:2013edk,Coutinho:2013lta,Queiroz:2016gif,Corcella:2017dns,Lindner:2016bgg,Nepomuceno:2019eaz}. We assess the relevance of these new decay channels for several benchmark models. 
%
%Furthermore, we forecast limits based on the High-Luminosity (HL), High-Energy (HE), and %Future Circular Collider (FCC-hh) LHC setups. 
Furthermore, under the assumption that no positive signal is found, we forecast limits for the High-Luminosity (HL) and High-Energy (HE) LHC setups, HL-LHC and HE-LHC, respectively, as well as for the Future Circular Collider (FCC-hh). Lastly, in the context of the 3-3-1 LHN model, we investigate if one can host a viable dark matter candidate in light of those bounds.\newline

In summary, our present  work expands previous studies in the following directions:

\begin{enumerate}
    \item taking into account updated data from LHC,
    \item considering 
   HL-LHC, HE-LHC and FCC setups
    %High-Luminosity and High-Energy LHC setups,
    \item contemplating overlooked $Z^\prime$ decays,
    \item connecting our findings with dark matter phenomenology.
\end{enumerate}

This work is organized as follows: in {\it Section} \ref{331M} we review the 3-3-1 model and the  relevant $Z^{\prime}$ decay channels; in {\it Section} \ref{DS} we discuss the benchmark models and kinematic cuts used in the production of the $Z^\prime$ signal; in {\it Section} \ref{R} we present our collider findings; in {\it Section} \ref{DMsec} we connect our collider results with dark matter phenomenology; and finally in {\it Section} \ref{dc} we draw our conclusions. 
\newline

\section{\label{331M} 3-3-1 RHN and LHN Models}

\subsection{Fermion Content}
In our work we analyze two models based on the local symmetry group $\textbf{SU(3)}_\textbf{C} \times \textbf{SU(3)}_\textbf{L} \times \textbf{U(1)}_\textbf{X}$, namely 3-3-1 with right-handed neutrinos  (3-3-1 RHN), \cite{Hoang:1996gi,Hoang:1995vq}, and 3-3-1 with neutral left-handed fermion (3-3-1 LHN) \cite{Mizukoshi:2010ky,Catano:2012kw}.
%,  where C represents color charge, L Left-handed, and X is the $\text{U(1)}_{X}$ quantum %number. 
The electric charge operator in these two models is the same, 
\begin{equation}
\frac{Q}{e}=\frac{1}{2}\left(\lambda_{3}-\frac{1}{\sqrt{3}}\lambda_{8}\right)+\text{X}\cdot\hat{\rm I},
\label{charged}
\end{equation}
where $\lambda_{3,8}$, are the diagonal generators of $\text{SU(3)}_\text{L}$ and $\hat{\rm I}$ is the identity matrix that acts as a generator of $\text{U(1)}_\text{X}$ group with $X$ being its corresponding charge.

The \textbf{3-3-1 RHN} model contains triplet and singlet fermionic fields 
with the following $\textbf{SU(3)}_\textbf{C} \times \textbf{SU(3)}_\textbf{L} \times \textbf{U(1)}_\textbf{X}$ assignments:
%Regarding the \textbf{3-3-1 RHN} model, it contains triplet and singlet fermionic fields %given by,
%
\begin{align}
&f_{aL}=\left(\begin{array}{c}
\nu_{L}^{a} \\
\ell_{L}^{a} \\
\nu_{R}^{a c}
\end{array}\right)_{} \sim(1,3,-1/3), \ \ \ell_{aR} \sim(1,1,-1), \label{fer}\\ \nonumber \\
&
Q_{i L}=\left(\begin{array}{c}
d_{i} \\
-u_{i} \\
d_{i}^{\prime}
\end{array}\right)_{L} \sim(3, \overline{3}, 0), \text{ } u_{i R} \sim(3,1,2 / 3), \\
&\hspace{0,8cm} d_{i R} \sim(3,1,-1 / 3),  \text{ } d_{i R}^{\prime} \sim(3,1,-1 / 3), \nonumber \\\nonumber \\
& 
Q_{3 L}=\left(\begin{array}{c}
u_{3} \\
d_{3} \\
T
\end{array}\right)_{L} \sim(3,3,1 / 3), \text{ } u_{3 R} \sim(3,1,2 / 3), \\
&\hspace{0,8cm} d_{3 R} \sim(3,1,-1 / 3), \text{ }T_{R} \sim(3,1,2 / 3), \nonumber
\end{align} where $a=1,2,3$ and $i=1,2$ indicate the generation indices. Notice that we have three new exotic quarks ${q}^{\prime}$ ($d_{i}^{\prime}$ and $T$). 

In the \textbf{3-3-1 LHN} a new heavy neutral lepton $N^{a}_{L}$ replaces the 
$\left( {\nu}_{R}^{a} \right)^{c}$ in the lepton triplet. Besides, a right-handed neutral fermion ${N}^{a}_R$ is introduced,  transforming as a singlet under ${SU(3)}_{L}$, 

\begin{equation}
    N^{a}_{R} \sim (1,1,0),
\end{equation} but the quark sector remains unchanged.

\begin{table}[t]
\begin{footnotesize}
\begin{center}
\begin{tabular}{|c|c|c|}
\hline
\multicolumn{3}{|c|}{$Z^{\prime}$ Interactions in the 3-3-1 model } \\
\hline
Interaction &  $g^\prime_V$ & $g^\prime_A$   \\ 

\hline
$Z^{\prime}\ \bar u u,\bar c c  $ &
$\displaystyle{\frac{3-8\sin^2\theta_W}{{6\sqrt{3-4\sin^2\theta_W}}}}$  & 
$\displaystyle{-\frac{1}{2\sqrt{3-4\sin^2\theta_W}}}$  \\
\hline
$Z^{\prime}\ \bar t t$ & 
$\displaystyle{\frac{3+2\sin^2\theta_W}{{6\sqrt{3-4\sin^2\theta_W}}}}$  & 
$\displaystyle{-\frac{1-2\sin^2\theta_W}{2\sqrt{3-4\sin^2\theta_W}}}$  \\
\hline
$Z^{\prime}\ \bar d d,\bar s s  $ &
$\displaystyle{\frac{3- 2\sin^2\theta_W}{6\sqrt{3-4\sin^2\theta_W}}}$  & 
$\displaystyle{-\frac{{3-6\sin^2\theta_W}}{6\sqrt{3-4\sin^2\theta_W}}}$  \\
\hline
$Z^{\prime}\ \bar b b$ & 
$\displaystyle{\frac{3-4\sin^2\theta_W}{{6\sqrt{3-4\sin^2\theta_W}}}}$  & 
$\displaystyle{-\frac{1}{2\sqrt{3-4\sin^2\theta_W}}}$  \\
\hline
$Z^{\prime}\ \bar \ell \ell $ &
$\displaystyle{\frac{-1+4\sin^2\theta_W}{2\sqrt{3-4\sin^2\theta_W}}}$ &
$\displaystyle{\frac{1}{2\sqrt{3-4\sin^2\theta_W}}}$ \\
\hline
$Z^{\prime} \overline{N} N $ &
$\displaystyle{\frac{4\sqrt{3-4\sin^2\theta_W}}{9}}$ &
$\displaystyle{-\frac{4\sqrt{3-4\sin^2\theta_W}}{9}}$ \\
\hline
$Z^{\prime}\ \overline{\nu_{\ell}} \nu_{\ell} $ &
$\displaystyle{\frac{\sqrt{3-4\sin^2\theta_W}}{18}}$ &
$\displaystyle{-\frac{\sqrt{3-4\sin^2\theta_W}}{18}}$ \\ \hline
$Z^{\prime} \overline{d^{i}_{i}} d^{i}_{i}$&$-\frac{3-5\sin^2\theta_W}{3 \sqrt{3-4\sin^2\theta_W}}$&$\frac{1-\sin^2\theta_W}{\sqrt{3-4\sin^2\theta_W}} $\\ \hline
$Z^{\prime} \overline{T} T $&$\frac{3-7\sin^2\theta_W}{ 3\sqrt{3-4\sin^2\theta_W}} $&-$\frac{1-\sin^2\theta_W}{\sqrt{3-4\sin^2\theta_W}} $\\
\hline
\end{tabular}
\end{center}
\end{footnotesize}
\caption{Vector and Axial couplings of the ${Z^{\prime}}$ boson with fermions in the 3-3-1 RHN and LHN models. In the 3-3-1 RHN model there are no interactions with the heavy fermions N. Apart from that, the $Z^\prime$ interactions are precisely the same as the 3-3-1 LHN model.}
\label{cc}
\end{table}

\subsection{Scalar Sector}

Fermion masses are obtained through the introduction of three scalar triplets, which we denote as $\chi$, $\rho$ and $\eta$ \cite{Mizukoshi:2010ky}, 
\begin{eqnarray}
\quad \quad  \chi=\left(\begin{array}{c}
\chi^{0} \\
\chi^{-} \\
\chi^{\prime 0}
\end{array}\right) \sim(1,3,-1 / 3), \quad \langle\chi\rangle=\left(\begin{array}{c}
0 \\
0 \\
v_{\chi}
\end{array}\right), \nonumber\\  \nonumber\\ 
\rho=\left(\begin{array}{c}
\rho^{+} \\
\rho^{0} \\
\rho^{\prime +}
\end{array}\right) \sim(1,3,2 / 3), \quad \quad  \langle\rho\rangle=\left(\begin{array}{c}
0 \\
v_{\rho} \\
0
\end{array}\right),\\ \nonumber\\ 
\eta=\left(\begin{array}{c}
\eta^{0} \\
\eta^{-} \\
\eta^{\prime 0}
\end{array}\right) \sim(1,3,-1 / 3), \quad \langle\eta\rangle=\left(\begin{array}{c}
v_{\eta} \\
0 \\
0
\end{array}\right).\nonumber \label{scalars}
\end{eqnarray}
where $v_{\chi}$, $v_{\rho}$ and $v_{\eta}$ correspond to the vacuum expectation values (VEVs) defining a two-step spontaneous symmetry breaking (SSB)
 \begin{align*}
    \textbf{SU(3)}_\textbf{L} \times \textbf{U(1)}_\textbf{X}  \xrightarrow{\langle\chi\rangle} \textbf{SU(2)}_\textbf{L} \times \textbf{U(1)}_\textbf{Y}\xrightarrow{ \langle\eta\rangle,\langle\rho\rangle} \textbf{U(1)}_\textbf{Q}.
\end{align*}

They form the scalar potential,

\begin{eqnarray} V(\eta,\rho,\chi)&=&\mu_\chi^2 \chi^2 +\mu_\eta^2\eta^2
+\mu_\rho^2\rho^2+\lambda_1\chi^4 +\lambda_2\eta^4
+\lambda_3\rho^4+ \nonumber \\
&&\lambda_4(\chi^{\dagger}\chi)(\eta^{\dagger}\eta)
+\lambda_5(\chi^{\dagger}\chi)(\rho^{\dagger}\rho)+\lambda_6
(\eta^{\dagger}\eta)(\rho^{\dagger}\rho)+ \nonumber \\
&&\lambda_7(\chi^{\dagger}\eta)(\eta^{\dagger}\chi)
+\lambda_8(\chi^{\dagger}\rho)(\rho^{\dagger}\chi)+\lambda_9
(\eta^{\dagger}\rho)(\rho^{\dagger}\eta) \nonumber \\
&&-\frac{f}{\sqrt{2}}\epsilon^{ijk}\eta_i \rho_j \chi_k +\mbox{H.c}.
\label{potential}
\end{eqnarray}

We have assumed $f=v_\chi$, $\lambda_2=\lambda_3$, $\lambda_4=\lambda_5$ to simplify our analytical results, but our conclusions are based on precise numerical calculations, where these simplifying assumptions are not made. The CP-even scalars give rise to the mass eingenstates, $S_1$, $S_2$ and the Higgs boson, with the following masses,

\begin{eqnarray}
m_{S_1}^2 = \frac{v^2}{4} +2 \lambda_1 v_{\chi}^2,\nonumber\\
m_{S_2}^2= \left(v_{\chi}^2 + 2v^2(2\lambda_2-\lambda_6)\right)/2,\nonumber\\
m_{H}^2= v^2(2\lambda_2 +\lambda_6),
\label{masscpeven}
\end{eqnarray}
whereas only a pseudoscalar mass eingenstate survives, $P_1$, where

\begin{eqnarray}
m^{2}_{P_{1}} = \frac{1}{2}(v_{\chi^\prime}^{2}+\frac{v^{2}}{2}).
\label{massp1}
\end{eqnarray}

A complex neutral scalar $\phi$ which is a combination of $\chi^0$ and $\eta^{0\prime}$ arises, as well as two charged scalars $h_1$ and $h_2$ whose masses are found to be,
\begin{eqnarray}
m^{2}_{\phi} & = & \frac{(\lambda_{7} + \frac{1}{2} )}{2}[v^{2}+v_{\chi^\prime}^{2}],
\label{massphi}
\end{eqnarray}
\begin{eqnarray}
m^{2}_{h^{-}_{1}} & = & \frac{\lambda_{8}+\frac{1}{2} }{2}(v^{2}+v_{\chi^\prime}^{2})\,, \nonumber \\
m^{2}_{h^{-}_{2}} & = & \frac{v_{\chi^\prime}^{2}}{2}+\lambda_{9}v^{2}\,.
\label{massash1h2}
\end{eqnarray}

These scalars are not relevant to our reasoning, but to clearly show this we will need Eqs.\eqref{masscpeven}-\eqref{massash1h2} and the gauge boson masses that we will cover below. 

\begin{table}[ht!]
\centering
 \caption{Implemented benchmark sets (BMs) corresponding to mass values of the heavy exotic quarks $q^{\prime}$ and the heavy neutral lepton $N$ in the 3-3-1 RHN and LHN models.}
\begin{tabular}{|c||c|c|c|}
\hline Model & \multicolumn{2}{|c|}{ 3-3-1 LHN } & 3-3-1 RHN \\ \hline
\hline Mass  & $M_{q^{\prime}} $ [TeV]& $M_{N}$ [TeV] & $M_{q^{\prime}}$ [TeV]\\  \hline
\hline BM1   & $10 $ & $10 $ & $10 $ \\
\hline BM2   & $1 $ & $10 $ & $1 $ \\
\hline BM3   & $1.5 $ & $10 $ & $1.5 $ \\
\hline BM4   & $2 $ & $10 $ & $2 $ \\
\hline BM5   & $2 $ & $2 $ & N/A \\
\hline BM6   & $2 $ & $2.5$ & N/A \\
\hline BM7   & $2 $ & $4 $ & N/A \\
\hline BM8   & $1$  & $1$  & N/A \\
\hline BM9   & $0.5$  & $10$  & N/A \\
\hline BM10   & $10$  & $0.5$  & N/A \\
\hline
\end{tabular}
    \label{all_BM}
\end{table}

\subsection{Gauge Bosons}

Throughout, we adopt the decoupling limit where the energy scale of SSB of the 3-3-1 symmetry is much larger than the electroweak one, i.e., $v_{\chi} \gg v_{\eta},v_{\rho}$.  As a result of the enlarged gauge group, new gauge bosons arise: $W^{\prime \pm}$, $U^{0}$, and a $Z^{\prime}$. Their masses are given by,

\begin{equation}
    \begin{aligned}
    m_{Z^{\prime}}^{2}&=\frac{g^{2}}{\left(3-4 s_{W}^{2}\right)}\left(c_{W}^{2} v_{\chi}^{2}+\frac{v_{\rho}^{2}+v_{\eta}^{2}\left(1-2 s_{W}^{2}\right)^{2}}{4 c_{W}^{2} }\right) ,\\
    m_{W^{\prime}}^{2}&=\frac{g^{2}}{4}\left(v_{\eta}^{2}+v_{\chi}^{2}\right),\quad 
    m_{U^{0}}^{2}=\frac{g^{2}}{4}\left(v_{\rho}^{2}+v_{\chi}^{2}\right),
    \end{aligned}
    \label{eqBosons}
\end{equation}where $ v^2 = v_{\eta }^{2}+v_{\rho}^{2} \simeq 246$ GeV, $g$ is the $SU(2)_{L}$ gauge coupling, ${c}_{W} \equiv cos\theta_{W}$, ${s}_{W} \equiv sin\theta_{W}$, with $\theta_{W}$ being the Weinberg angle.  From Eq.~\eqref{eqBosons}, one can clearly see that the gauge boson masses are determined by $v_\chi$. Hence, once we set a bound on the $Z^\prime$ mass, it can be translated into a constraint on the $W^\prime$ and $U^0$ masses as well. One should notice that $W^\prime$, $Z^\prime$ and $U^0$ bosons have similar masses. 

We derive the limit on the $Z^\prime$ mass using the high-mass dilepton resonance searches at the ATLAS detector with $\sqrt{s}=13$ TeV center-of-mass-energy, and later estimate the future collider bounds. In that regard, the main ingredient is the neutral current that reads
\begin{equation}  \label{eq7}
		\mathcal{L}^{N C}_{ Z^{\prime} f f } = \frac{g}{2 c_{W}}\bar{f} \gamma^{\mu}\left[\text{g}_{\text{V}}^{(f)}+\gamma_{5}\text{g}_{\text{A}}^{(f)}\right] f Z_{\mu}^{\prime},
\end{equation}
where $\text{g}_{\text{V}}^{(f)}$ and ($\text{g}_{\text{A}}^{(f)}$) are the vector (axial) coupling constant of fermions $f = \ell,{N},{q}^{\prime}$ with $Z^{\prime}$ (see \tab{cc}). % and ${c}_{W} = cos\theta_{W}$, where $\theta_{W}$ is the Weinberg angle. 
The branching ratio of the $Z^{\prime}$ boson in two charged leptons is defined as
\begin{equation}
\operatorname{Br}\left(Z^{\prime} \rightarrow \ell \bar{\ell}\right)=\frac{\Gamma\left(Z^{\prime} \rightarrow \ell \bar{\ell}\right)}{\Gamma_{Z^{\prime}}},
\end{equation}
where $\Gamma_{Z^{\prime}}$ is the total width %{\color{blue} \st{define as} }
\begin{eqnarray}
\label{eq:Z'-Gamma}
\Gamma_{Z^{\prime}} =\sum_{X}\Gamma\left(Z^{\prime} \rightarrow 2X \right), 
\end{eqnarray}
being $X$ the SM particles and new particles in 3-3-1 models. The $\Gamma\left(Z^{\prime} \rightarrow \ell \bar{\ell}\right)$ is the partial decay width into dileptons at leading order, with  $\ell=e,\mu$,
\subsection{Importance of Scalars and Gauge Boson Decays decays}

The decay widths were actually computed using Calchep~\cite{CALCHEP}. As we pointed out before, our calculation of the total width takes into account all possible decays, including new gauge bosons, scalars, exotic quarks and dark matter whenever they are allowed. 

We have written explicitly the scalar masses to address the relevance of scalar fields to our reasoning. Bear in mind $m_{Z^\prime} \simeq 0.3 v_\chi$, and the relevant  $Z^\prime$ interactions are: $Z^\prime \phi \phi^{\ast}$, $Z^\prime W^\prime h_1^-$, $Z^\prime h_1^- h_1^+$,$Z^\prime h_2^- h_2^+$, $Z^\prime W^+ h_2^-$, $Z^\prime P_1 S_1$, among others. Looking at Eq.\ref{masscpeven}-\ref{massash1h2} it is clear that the scalars are much more massive than the $Z^\prime$ gauge boson, and thus do not contribute to the two-body $Z^\prime$ decay width. Three-body decays widths are possible, but suppressed. For this reason, we can solidly state that scalars do not play a role in our phenomenology. Moreover, as the exotic gauge bosons have similar masses, $Z^\prime$ decays into exotic boson pairs are not kinematically accessible. There are exotic decays into dark matter and exotic quarks that are important, however, but we will address in the next section. 
Furthermore, as the scalar are much more massive than the $Z^\prime$ gauge boson, they are not within reach LHC. In summary, scalar fields do not offer a possible signature for a 3-3-1 symmetry at the LHC.

\section{\label{DS} Data and signal output}

We carry out our collider simulation with Madgraph5 \cite{alwall2014automated,frederix2018automation}, and compute the decay with CalcHEP \cite{belyaev2013calchep,CALCHEP}. We compute the $pp \to Z^{\prime} \to \ell\bar{\ell}$ at  $\sqrt{s}=13$ TeV, with $\ell = e,\mu$ and compare our findings with the public results from ATLAS Collaboration \cite{atlas2019search}. We generate the Monte Carlo events to simulate the cross-section of the  Drell-Yan process using the parton distribution function (PDF) NNPDF23LO \cite{Carrazza:2013axa}. We require two opposite charge leptons in the event and the following kinematic cuts in order to compare our results with the ATLAS Collaboration data:  $p_{\mathrm{T}}>30$ GeV\footnote{Transverse momentum is the component of the momentum that is perpendicular to the beam axis.}  and $|\eta|<2.5$ \footnote{The pseudo-rapidity $\eta$ is defined as, $\eta=-\ln \tan \frac{\theta}{2}$, where $\theta$ is the polar angle between the particle's linear momentum  and the positive direction of the beam axis.}. 

Instead of considering only $Z^\prime$ interactions to fermions as done in previous works, we fully implemented the model in LANHEP \cite{semenov1998lanhep,semenov2009lanhep,semenov2016lanhep}, SARAH-HEP \cite{SARAH, vicente2017computer,staub2015exploring} and Feynrules \cite{alloul2014feynrules,Feynrules} to generate the output files for CalcHeP and Madgraph5, respectively.
This is important because additional exotic decays of the $Z^\prime$ gauge boson can significantly weaken the lower mass bounds based on dilepton data. The more sizeable decay channels are added to the total width, the smaller is the branching ratio into dileptons. Consequently, weaker limits are found. We investigate the importance of each of the possible new decay channels by considering several benchmark models.

\subsection{\label{IIIA} Importance of exotic $Z^{\prime}$ decays and benchmarks models}

In the 3-3-1 models explored here, the $Z^\prime$ might decay into SM fermions, new scalars, new gauge bosons ($W^{\prime \pm}$, $U^0, U^{0 \dagger}$), and invisibly ($N_i$). One of the three heavy fermions is cosmologically stable, and is rendered as dark matter candidate. The other two are long-lived. Suppose $N_1$ is the lightest one, for the sake of the argument. Because of a $Z_2$ symmetry where $N_i \rightarrow -N_i$, $N_2$ might decay into $N_1$ via the $W^\prime$ gauge boson, but this decay width is suppressed for two reasons: the $W^\prime$ is heavy, and second, the entries of the mixing matrix involving $N_1-N_2-N_3$ should also be small otherwise one could observe lepton flavor violation processes as explored in \cite{Arcadi:2017xbo}. As far as collider searches are concerned, without worrying about particular details of the masses and mixing matrices, we can safely take the neutral fermions as stable particle, and thus rendered as missing energy. 

We have checked that decays into new scalars and exotic gauge bosons are either very suppressed or kinematically prohibited for the benchmark points of Table~\ref{all_BM}. Hence, the only relevant new decay channels beyond the SM, are those involving exotic quarks, and neutral fermions ($N_i$). 

Hence, we investigate several benchmark models varying the masses of the decay products to quantify their importance in the derivation of lower mass bounds.

\subsection{Methods}

We compute the $ \sigma \times  \operatorname{BR}(\ell  \overline{\ell})$ at the LHC using the aforementioned  high-energy physics tools for each benchmark model (BM) in Table \ref{all_BM}, and later compare our results with ATLAS data.  Furthermore, we use these results to obtain, new bounds for HL-LHC, HE-LHC and FCC-hh colliders. 
To this end we apply Collider Reach  ($\beta$) tool, which takes the input bound on $m_{Z^{\prime}}$ obtained in the first
step at a certain center-of-mass-energy and luminosity, and forecasts new bounds for a different collider configurations including center-of-mass-energy and luminosity. In our work, we are interested in the High-Luminosity, High-Energy \cite{Cepeda:2019klc}, and the FCC proton-proton setups \cite{FCC:2018byv}.

\section{\label{R}Results and discussions}

\paragraph{\label{gbr} \textbf{Branching ratios.}}  

The partial widths of the $Z^\prime$ into charged leptons is the same for both 3-3-1 models, as they have identical interactions (see \eq{eq7}, and \tab{cc}), yet the branching ratio into charged leptons can be quite different for these two models. The key difference of the 3-3-1 LHN from the 3-3-1 RHN is the presence of heavy neutral fermions, $N_i$. Only when decays into $N_i$ pairs are inaccessible, both models are indistinguishable as far as $Z^\prime\to \ell^+\ell^-$ searches are concerned.
As aforementioned, section \ref{IIIA}, the scalars are not relevant to our results. Despite decays to scalars are present in some benchmarks, they are negligible compared to decays into SM quarks and leptons.

\begin{figure}[ht!]
    \centering
    \includegraphics[width=0.8\columnwidth]{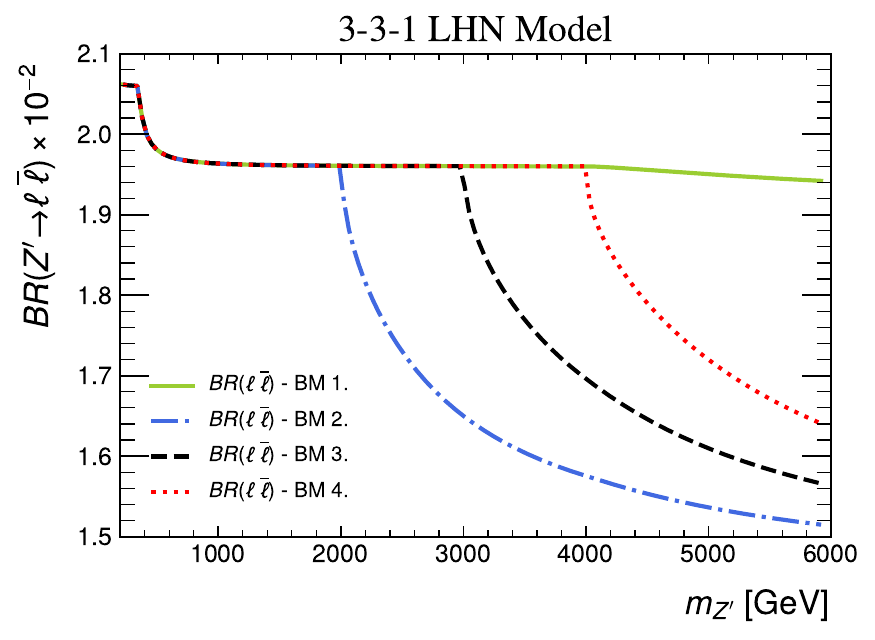}
    \caption{ Branching ratio for the $Z^{\prime}$ decay into dilepton channel as a function to $m_{Z^{\prime}}$ for the benchmark sets BM1, BM2, BM3, and BM4 of the 3-3-1 LHN model. }
    \label{BrLHN1}
\end{figure}

\begin{figure}[ht!]
    \centering    
    \includegraphics[width=0.8\columnwidth]{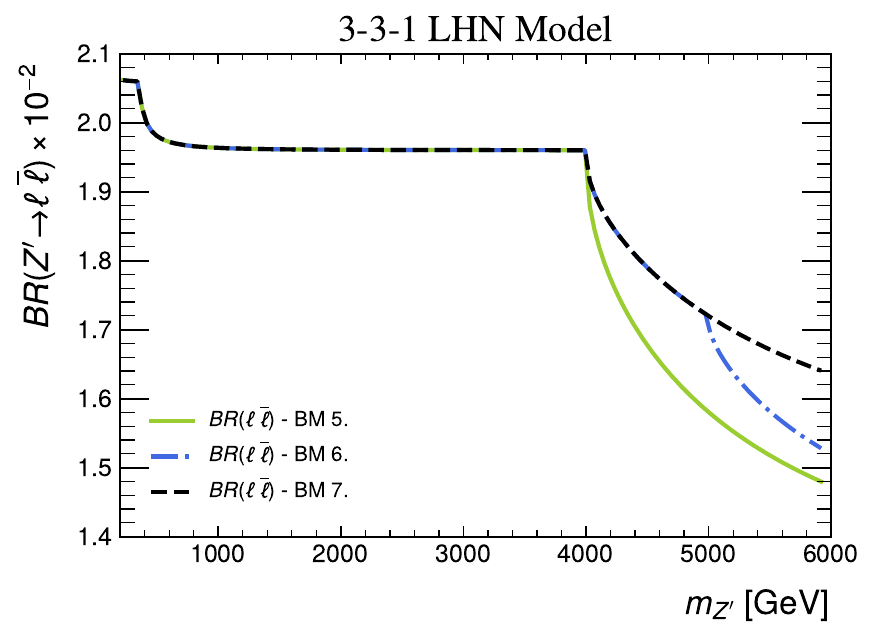}
    \caption{ Branching ratio for the $Z^{\prime}$ decay into dilepton channel as a function of $m_{Z^{\prime}}$ for the benchmark sets BM 5, 6 and 7 of the the 3-3-1 LHN model.}
    \label{BrLHN2}
\end{figure}

\begin{figure}[ht!]
    \centering
    \includegraphics[width=0.8\columnwidth]{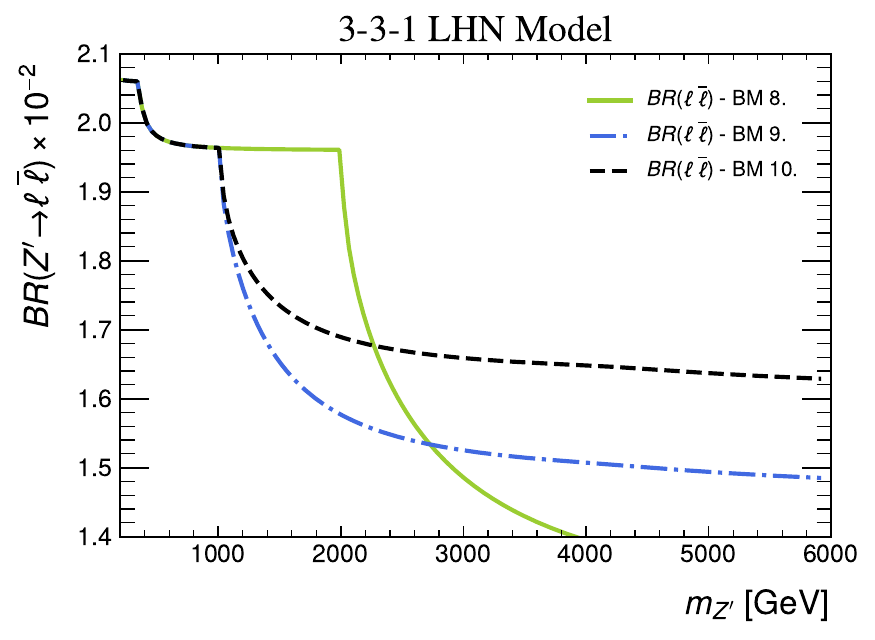}
    \caption{Branching ratio for the $Z^{\prime}$ decay into dilepton channel as a function to $m_{Z^{\prime}}$ for the benchmark sets BM 8, 9, and 10 the 3-3-1 LHN model.}
   \label{BrLHN3}
\end{figure}

\begin{figure}[ht!]
    \centering
    \includegraphics[width=0.8\columnwidth]{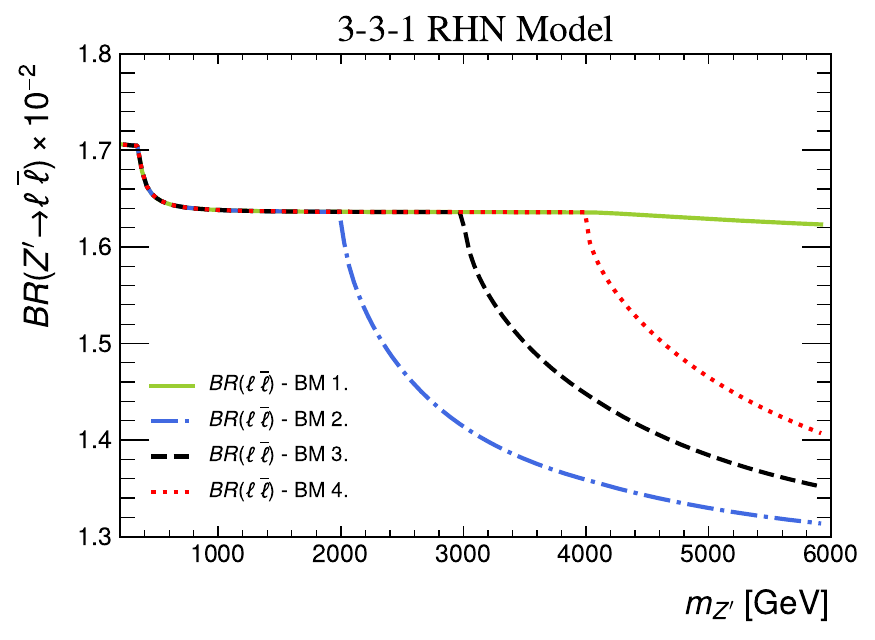}
    \caption{Branching ratio for the $Z^{\prime}$ decay into dilepton channel as a function to $m_{Z^{\prime}}$ for the 3-3-1 RHN model.}
    \label{BrRHN}
\end{figure}

We have shown in \figs{BrLHN1}, \ref{BrLHN2}, \ref{BrLHN3}, and \ref{BrRHN} the branching ratio $BR(Z^{\prime} \to \ell\bar{\ell})$ as a function of $m_{Z^{\prime}}$  for several BM in the 3-3-1 RHN and LHN models. We notice that in both models, the value of the branching ratio is less than 2\% and 1.7\%, respectively. In \fig{BrLHN1}, for the 3-3-1 LHN, the BM 2-3-4 lead to a drop near $2000$~GeV, $3000$~GeV, and $4000$~GeV, respectively. This behavior is caused by $Z^\prime$ decay into exotic quark. In contrast, the BM1 does not experience such behavior since the new exotic quarks masses are fixed at 10 TeV. 

\begin{table}[ht]
\centering
\caption{$m_{Z^{\prime}}$ lower bounds taking into account the dilepton signal data at the LHC \cite{atlas2019search} and the theoretical signal production from \fig{331LHN_all_BMs} for the 3-3-1 RHN and LHN models. }
    \begin{tabular}{|c||c|c|}
    \hline Model  & BM   & $m_{Z^{\prime}}$[GeV] \\ \hline \hline
        \multirow{5}{2cm}[1 mm]{3-3-1 RHN} 
                    & BM 1 \footnote{The lower bounds of BM 1 for the 3-3-1 RHN model are equivalent to those of BM 10 in the 3-3-1 LHN model.}     
                                      & $4052$   \\\cline{2-3}
                    & BM 2            & $3960$   \\\cline{2-3}
                    & BM 3 \footnote{The lower bound of BM 3 for the 3-3-1 RHN model i  equivalent to those of BM 9 in the 3-3-1 LHN model.}
                                      & $3989$   \\\cline{2-3}
                    & BM 4            & $4040$   \\\cline{1-3}
        \multirow{9}{2cm}[1 mm]{3-3-1 LHN }
                    &  BM 1           & $4132$   \\\cline{2-3}
                    &  BM 2           & $4013$   \\\cline{2-3}
                    &  BM 3           & $4060$   \\\cline{2-3}  
                    &  BM 4, 6 and 7  & $4118$   \\\cline{2-3}
                    &  BM 5           & $4094$   \\\cline{2-3}
                    &  BM 8           & $3950$   \\ \hline
    \end{tabular}
    \label{bounds}
\end{table}

\begin{figure*}[!ht]
    \centering
     \subfigure[$\sigma_{\text {fid.}} \times \operatorname{BR}(\ell \bar{\ell})$ vs $m_{Z^{\prime}}$ for BM sets 1, 2, and 3 in the 3-3-1 RHN model. \label{xsec0} ]{\includegraphics[scale=0.45]{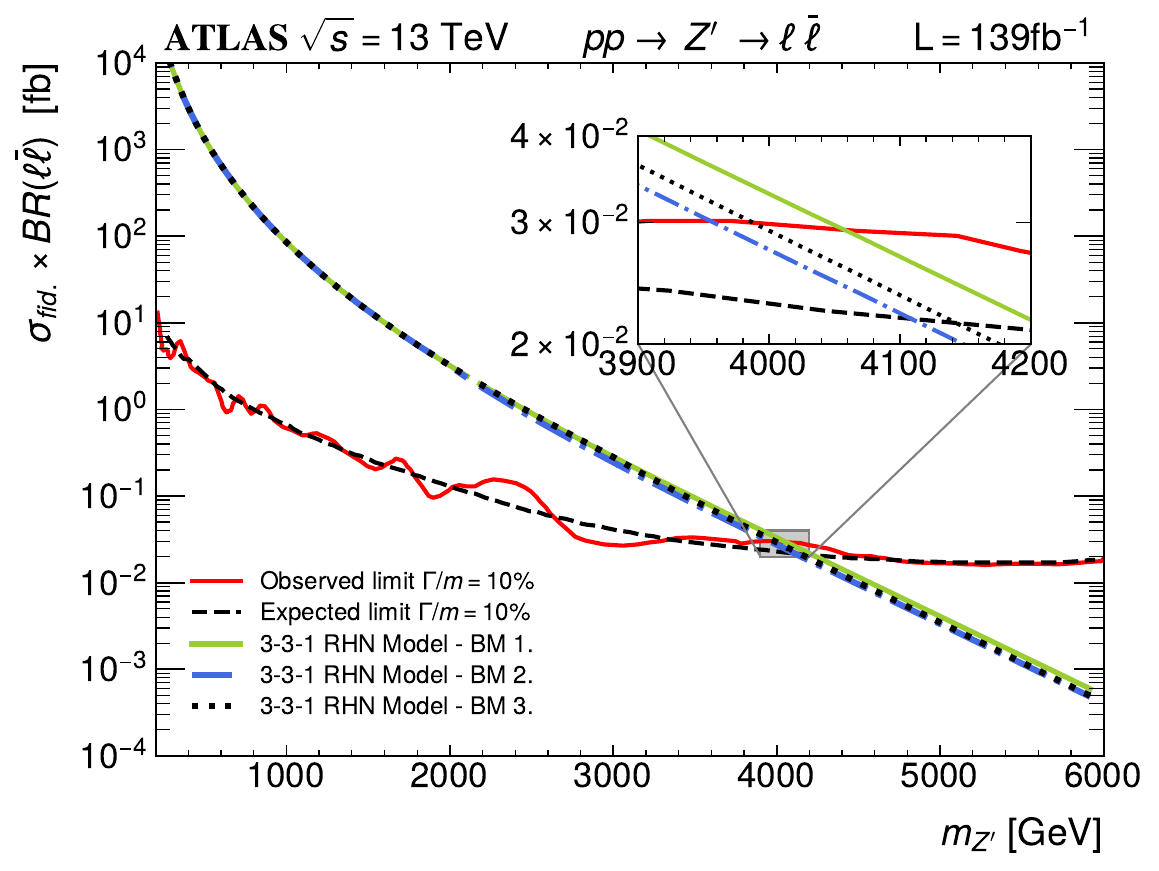}}
    \subfigure[$\sigma_{\text {fid.}} \times \operatorname{BR}(\ell \bar{\ell})$ vs $m_{Z^{\prime}}$ for BM sets 1, 2, and 3 in the 3-3-1 LHN model. \label{xsec1} ]{\includegraphics[scale=0.45]{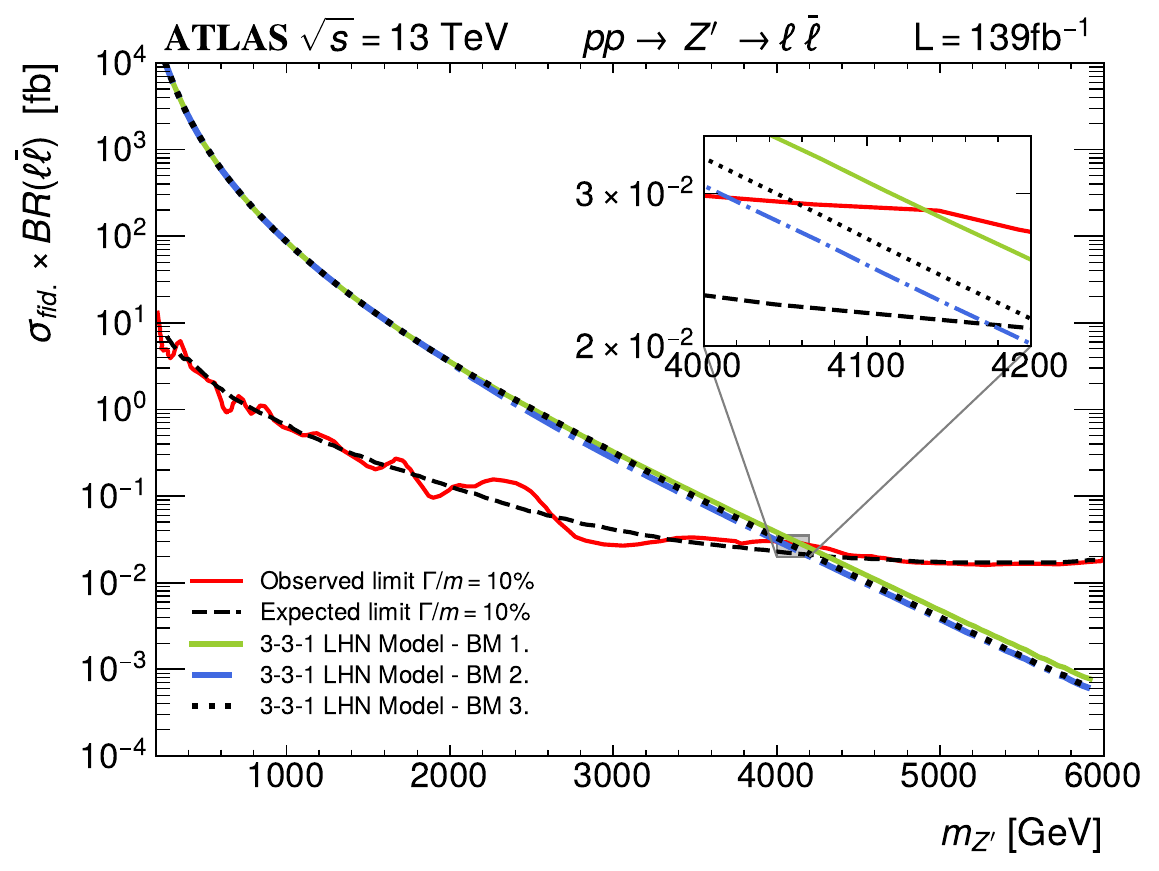}}
    \subfigure[$\sigma_{\text {fid.}} \times \operatorname{BR}(\ell \bar{\ell})$ vs $m_{Z^{\prime}}$ for BM sets 5, 6 and 7 in the 3-3-1 LHN model.\label{xsec2} ]{\includegraphics[scale=0.45]{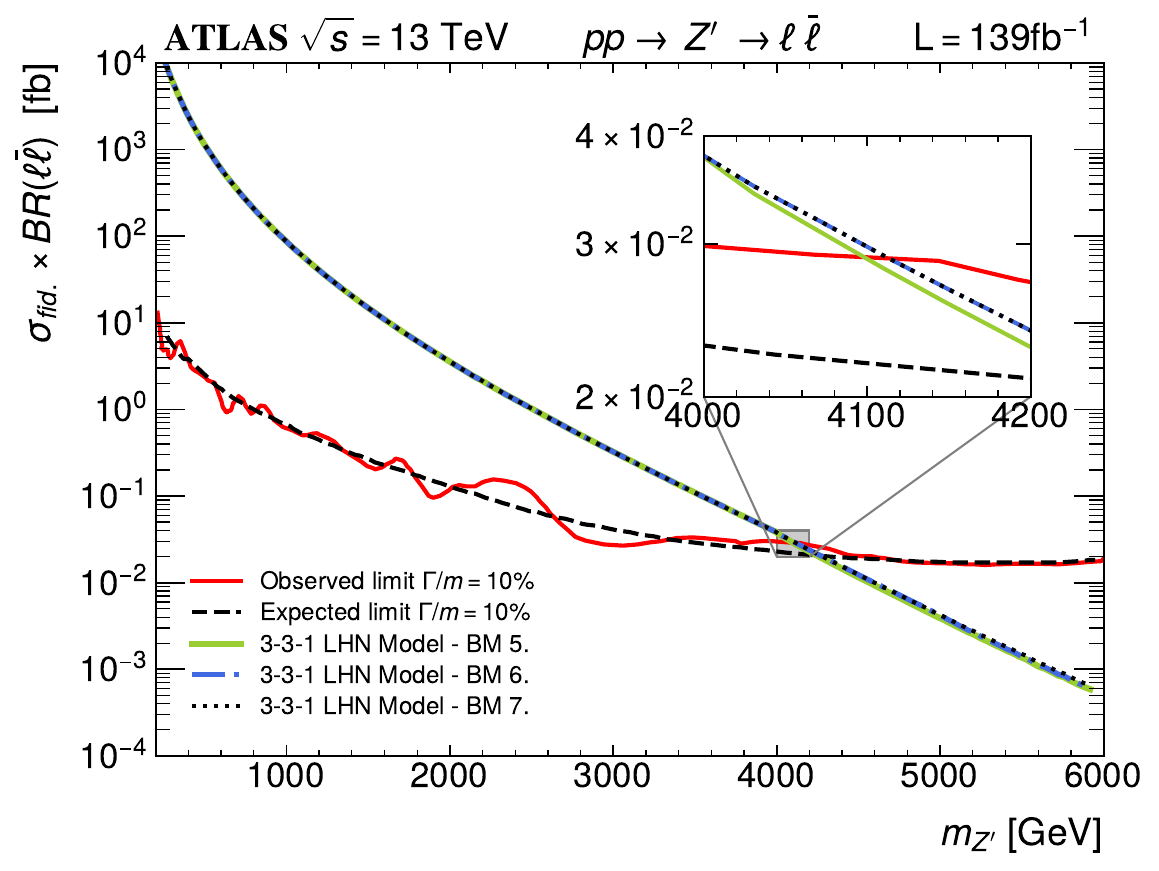}}
    \subfigure[$\sigma_{\text {fid.}} \times \operatorname{BR}(\ell \bar{\ell})$ vs $m_{Z^{\prime}}$ for BM sets 8, 9 and 10 in the 3-3-1 LHN model. \label{xsec3} ]{\includegraphics[scale=0.45]{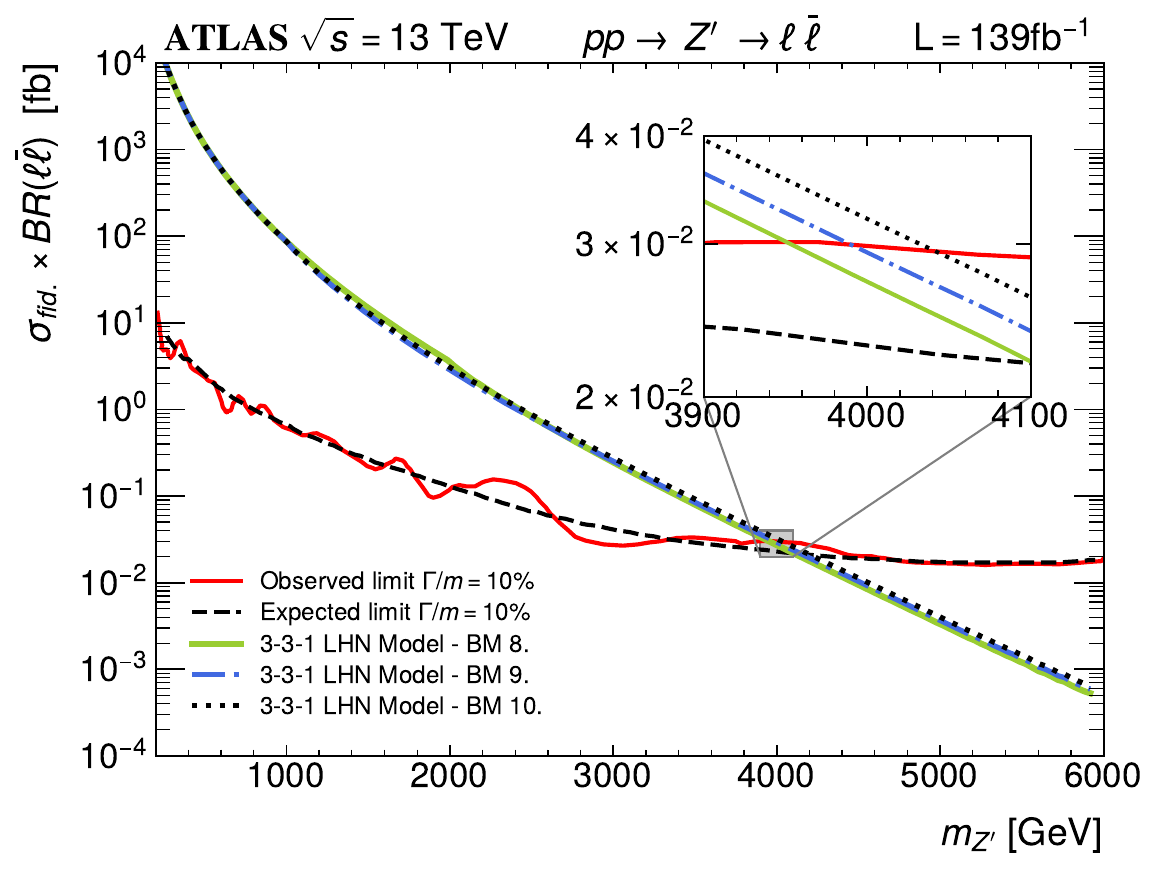}}
    \caption{Solid red and dashed black lines symbolize $\sigma_{\text {fid.}} \times \operatorname{BR}(\ell \bar{\ell})$ upper limits observed and expected at $95 \% \mathrm{CL}$ as a function of $Z^{\prime}$ mass for the $10 \%$ width signals for the dilepton channel $Z^{\prime} \rightarrow \ell \bar{\ell}$ in the ATLAS experiment at a center of mass energy 13 TeV (ATLAS Collaboration \cite{atlas2019search}). The solid yellowgreen, dash-dot blue, and black dotted lines represent the theoretical production $\sigma{(pp \to Z^{\prime})}\times BR{(Z^{\prime} \to \ell\bar{\ell})}$ generated using MadGraph5 and CalcHEP for several benchmark sets for the 3-3-1 RHN and LHN models. We assume different masses for the new exotic quarks and heavy neutral lepton (see \tab{all_BM}). The lower mass bounds on the $Z^{\prime}$ obtained can be seen in \tab{bounds}.}
\label{331LHN_all_BMs}
\end{figure*}

In the same way, for the 3-3-1 LHN model, we observed some substantial deacrease in the branching ratio into charged leptons for the BM sets 5, 6, and 7 when $m_{Z^\prime} = 4000$~ GeV (see \fig{BrLHN2}). It has to do with the exotic quarks at $m_{q^{\prime}} = 2000$~GeV. The importance of decays into heavy neutral fermions can be seen in BM 6, which leads to a significant decrease in the branching ratio when $m_{Z^\prime} \sim 5$~TeV, as we fixed $m_{Ni} =2.5$~TeV. In a similar vein, the behavior seen in \fig{BrLHN3} can be explained.

For the 3-3-1 RHN model, we observe that the behavior for the branching ratio is similar to the 3-3-1 LHN model since by performing the same variations on the exotic quark masses, we have the same decrease in the branching ratio (see \fig{BrRHN}). However, the size of the branching ratio into charged leptons is smaller because the $Z^\prime$ can always decay into right-handed neutrinos, which are assumed to have keV masses \cite{Dias:2005yh}.

\paragraph{ \bf{Signal production} \label{xsec}} 

As explained above, the theoretical production of an $Z^{\prime}$ at the LHC decaying into dileptons was generated using MadGraph5 and CalcHEP. To compare the theoretical signal for the dilepton channel $Z^{\prime}$ with ATLAS collaboration data presented in Fig.~3(a) in \cite{atlas2019search}, we plot $\sigma{(pp \to Z^{\prime})}\times BR{(Z^{\prime} \to \ell\bar{\ell})}$ as a function of $m_{Z^{\prime}}$ for the 3-3-1 RHN and LHN models,  as seen in \fig{331LHN_all_BMs}. For the $Z^{\prime}$ mass we take different values in the interval of 200 GeV$<m_{Z^{\prime}}<$6000 GeV with steps of $40$ GeV. The lower mass bounds on the $Z^{\prime}$ are obtained by considering the intersection of the solid yellowgreen, dash-dot blue, and black dotted lines with the red solid curve in \figs{xsec0}, \ref{xsec1}, \ref{xsec2} and \ref{xsec3}, and these results are summarized in \tab{bounds}.

\paragraph{ \bf{HE-HL and FCC-hh colliders.} \label{HEL}}

After obtaining the lower bounds of ${m}_{ {Z}^{\prime} }$ for the 13 TeV LHC after 139 fb$^{-1}$ (\tab{bounds}), we use these results as input for Collider Reach (${\beta}$) with the PDF \texttt{MMHTMMHT2014nnlo68cl}~ \cite{Harland-Lang:2014zoa}, and obtain the expected limits for HL-LHC, HE-LHC and FCC-hh setups. 

We set the following collider configurations:
\begin{itemize}
\item HE-HL : for the center-of-mass energy $\sqrt{s} = 13$~TeV, $14$~TeV and 27~TeV, and integral luminosity $L_{int}= $ 139 fb$^{-1}$, 300 fb$^{-1}$, 500 fb$^{-1}$, and 3000 fb$^{-1}$.
\item FCC-hh: for the center-of-mass energy $\sqrt{s} = 100$~TeV and integral luminosity $L_{int}= $ 139 fb$^{-1}$, 300 fb$^{-1}$, 500 fb$^{-1}$, and 3000 fb$^{-1}$.
\end{itemize}

The mass reach, are displayed in \tab{HE-HL-bounds} for HE-HL and FCC-hh collider. At the HL-LHC, the expected lower mass bounds raise by 1.2--1.5 TeV compared to the 139 fb$^{-1}$ data. \tab{bounds}. In special, with 3000 fb$^{-1}$ at the 14 TeV LHC, the projected sensitivity increases by almost 2 TeV compared to the current bounds for some benchmark points.

At the 27 TeV HE-LHC and the 100 TeV FCC-hh collider, with $L_{int}= $ 3000 fb$^{-1}$, the lower mass bounds improve by a factor of $\sim 2.5$ and $\sim 7$, respectively, compared to those obtained at the LHC with a center-of-mass energy $\sqrt{s} = 13$~TeV and integral luminosity of 139 fb$^{-1}$ (see \tab{bounds}).  
%Besides, we note that increasing the center-of-mass energy from $13$~TeV to $14$~TeV the mass ranges become closer. 

%\sout{These \sout{new bounds} \Yoxara{ or mass reach} are displayed in \tab{HE-HL-bounds}, for which we use different values for the center-of-mass energy $\sqrt{s} = 13$~TeV, $14$~TeV, $27$~TeV, and $100$~TeV, and for the integral luminosity $L_{int}= $ 139 fb$^{-1}$, 300 fb$^{-1}$, 500 fb$^{-1}$, and 3000 fb$^{-1}$, respectively.} 

We also note that BM 1 and 3, in the 3-3-1 RHN model, coincide with BM 10 and 9 in the 3-3-1 LHN model, respectively. Moreover, BM 6-7 present a similar bound to BM 4 in the 3-3-1 LHN model, and these results are easily justified by the presence or not of exotic $Z^\prime$ decays as discussed previously. 

Having in mind that 3-3-1 LHN features heavy neutral fermions that can be dark matter candidates, one may wonder if the current and projected collider bounds derived in our work preclude the existence of a plausible dark matter candidate in the model. We address this concern below.

%\begin{figure}[ht!]
 %   \centering
  %   \includegraphics[scale=0.6]{PLots/HEHL/HL_27TeV.pdf}
 %   \caption{Lower bounds of the $m_{Z^{\prime}}$ as a function of the integral luminosity $L_{int}$ with a center-of-mass energy of $\sqrt{s} = $ 27TeV for the  3-3-1  RHN and LHN models.}
 %   \label{HE_HL}
%\end{figure}

%\begin{figure}[ht!]
 %   \centering
  %   \includegraphics[scale=0.6]{PLots/HEHL/HL_100TeV.pdf}
  %  \caption{Lower bounds of the $m_{Z^{\prime}}$ as a function of the integral luminosity $L_{int}$ with a center-of-mass energy of $\sqrt{s} = $ 100TeV for the  3-3-1  RHN and LHN models.}
   % \label{FCC}
%\end{figure}

\section{\label{DMsec} dark matter}

 \begin{figure}[h!]
\centering
\includegraphics[width=0.9\columnwidth]{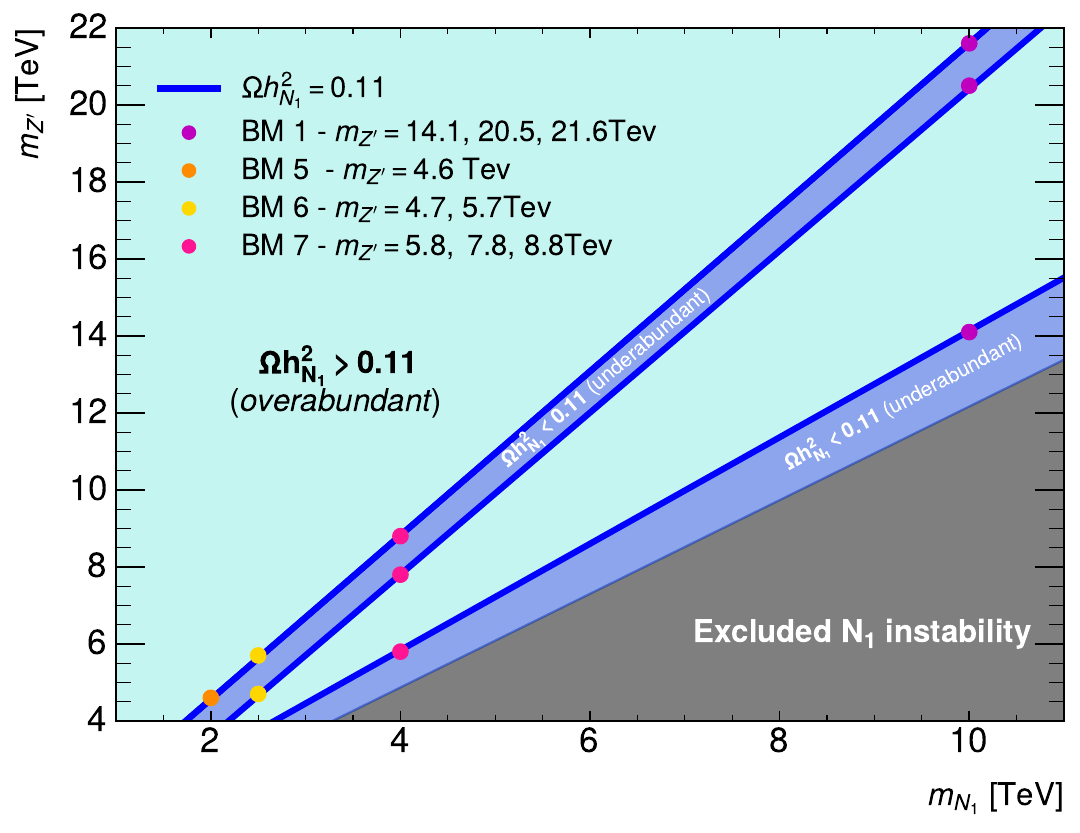}     \caption{Parameter space $m_{Z^{\prime}} \times m_{N_{1}}$ plane that explains the thermal relic density. BM models are indicated in the figure, as well the regions that lead to overabundant and underabundant dark matter. In the gray region, $N_1$ is not stable.}
  \label{DM1}
\end{figure}

\paragraph{ \bf{Thermal Production} \label{secDM1}}

The lightest of the neutral fermions, let us say $N_1$, can be a viable dark matter candidate due to a $Z_2$ symmetry \cite{Mizukoshi:2010ky,Profumo:2013sca}. The dark matter abundance is governed by s-channel annihilations into SM fermions mediated by the $Z^\prime$ gauge field. The $Z^\prime$ interactions with SM fermions are set by the gauge symmetry, which features a fixed gauge coupling. In other words, the dark matter abundance is governed by two parameters only, the dark matter and the $Z^\prime$ masses \cite{Mizukoshi:2010ky,Oliveira:2021gcw}. Furthermore, the dark matter scattering off nucleon occurs through a t-channel $Z^\prime$ exchange. Hence, the dark matter phenomenology is quite predictable once we fix the $Z^\prime$ mass.  

In the context of thermal production, where the production of dark matter occurs in the usual standard freeze-out, the curve that yields the correct relic density, $\Omega h^2=0.11$ \cite{Planck:2018vyg}, is shown in \fig{DM1} with solid blue curves. We also exhibit the region where the dark matter is overabundant and underabundant. Considering current LHC bound, we observe that for the BM 5, $m_N=2$~TeV, we can reproduce the correct relic density for $m_{Z^{\prime}} \sim 4.6$~TeV, which safely obeys LHC constraints. However, with $500 fb^{-1}$ of data, LHC will already be able to probe this scenario (see \tab{HE-HL-bounds}).
For the BM 6, the right relic density is found for $m_{Z^{\prime}} \sim 4.7$ and  $5.7$~TeV, whereas for BM7 a $5.8$~TeV, $7.8$~TeV and $8.8$~TeV $Z^\prime$ could yield the correct relic density. In particular, for the BM 1, only HE-LHC has the potential to fully probe this scenario, as it can exclude $Z^\prime$ masses up to $9.8$~TeV. There are many interesting scenarios to be explored, but one can solidly see the importance of orthogonal and complementary searches for new physics. Our study, clearly shows that setting aside one's theoretical prejudice for multi-TeV mediators, vanilla thermal dark matter models, such as the one present in the 3-3-1 LHN model, is fully consistent with current LHC data, and only the next generation of colliders will be able to close the few TeV dark matter particle window.

\paragraph{ \bf{Direct Detection} \label{secDM2}}

The neutral fermion can indeed leave signals at direct detection experiments through t-channel $Z^\prime$ exchange, but the limits are weak compared to those stemming from collider searches. This is a typical feature of vector mediator models that have sizeable couplings to fermions \cite{Alves:2013tqa,Alves:2015pea}. The current and projected direct detection bounds, from XENON or PANDAX collaborations \cite{XENON:2018voc,PandaX-4T:2021bab,XENON:2020kmp} are significantly surpassed by current LHC data, mainly when are put into perspective with future colliders. For this reason, we decided not to show them in \fig{DM1}.

\begin{table*}[ht!]
\centering
\caption{$m_{Z^{\prime}}$ mass reach for all benchmark sets considered in this work at HE-HL and FCC-hh colliders by increasing the center-of-mass energy ($\sqrt{s}$) from 13TeV until 100TeV, and integral luminosity ($L_{int}$) from 139 $fb^{-1}$ to 3000 $fb^{-1}$, for the 3-3-1 RHN and LHN models. Values of $m_{Z^{\prime}}$ for HE-HL LHC appear between the fourth and sixth columns of the table, whereas for the FCC-hh collider, the $m_{Z^{\prime}}$ reachs are shown in the seventh column, when increasing the luminosity (column three).}
    \begin{tabular}{|c||c|c|c|c|c|c|}
            \hline Model  & Benchmark (BM) sets & $L_{int}$[fb$^{-1}$]  & $m_{Z^{\prime}}$[TeV]-13TeV & $m_{Z^{\prime}}$[TeV]-14TeV    & $m_{Z^{\prime}}$[TeV]- 27TeV  & $m_{Z^{\prime}}$[TeV]-100TeV   \\ \hline \hline
        \multirow{21}{2cm}[3 mm]{3-3-1 RHN }
         &               & 139  & $4.052$ & $4.288$ & $6.987$ & $17.180$  \\\cline{3-7}
         &               & 300  & $4.390$ & $4.651$ & $7.675$ & $19.447$  \\\cline{3-7}
         & BM 1  \footnote{The lower bounds of BM 1 for the 3-3-1 RHN model are equivalent to those of BM 10 in the 3-3-1 LHN model.}
                         & 500  & $4.613$ & $4.892$ & $8.136$ & $21.006$  \\\cline{3-7}
         &               & 1000 & $4.916$ & $5.217$ & $8.763$ & $23.175$  \\\cline{3-7}
         &               & 3000 & $5.388$ & $5.727$ & $9.755$ & $26.711$  \\\cline{2-7}      
         &               & 139  & $3.960$ & $4.189$ & $6.801$ & $16.548$  \\\cline{3-7}
         &               & 300  & $4.298$ & $4.552$ & $7.487$ & $18.821$  \\\cline{3-7}
         & BM 2          & 500  & $4.521$ & $4.793$ & $7.947$ & $20.363$  \\\cline{3-7}
         &               & 1000 & $4.825$ & $5.119$ & $8.574$ & $22.514$  \\\cline{3-7}
         &               & 3000 & $5.298$ & $4.699$ & $9.566$ & $26.030$  \\ \cline{2-7}
         &               & 139  & $3.989$ & $4.220$ & $6.860$ & $16.769$  \\\cline{3-7}
         &               & 300  & $4.327$ & $4.583$ & $7.547$ & $19.016$  \\\cline{3-7}
         & BM 3 \footnote{The lower bounds of BM 3 for the 3-3-1 RHN model are equivalent to those of BM 9 in the 3-3-1 LHN model.}         
                         & 500  & $4.550$ & $4.824$ & $8.006$ & $20.564$  \\\cline{3-7}
         &               & 1000 & $4.853$ & $5.149$ & $8.633$ & $22.721$  \\\cline{3-7}
         &               & 3000 & $5.326$ & $5.661$ & $9.626$ & $26.244$  \\ \cline{2-7}
         &               & 139  & $4.040$ & $4.275$ & $6.963$ & $17.101$  \\\cline{3-7}
         &               & 300  & $4.378$ & $4.638$ & $7.651$ & $19.364$  \\\cline{3-7}
         & BM 4          & 500  & $4.601$ & $4.879$ & $8.111$ & $20.921$  \\\cline{3-7}
         &               & 1000 & $4.904$ & $5.204$ & $8.739$ & $23.089$  \\\cline{3-7}
         &               & 3000 & $5.377$ & $5.715$ & $9.731$ & $26.652$  \\ \hline \hline
        \multirow{31}{2cm}[3 mm]{3-3-1 LHN}
         &               & 139  & $4.132$ & $4.374$ & $7.149$ & $17.709$  \\\cline{3-7}
         &               & 300  & $4.470$ & $4.737$ & $7.839$ & $19.990$  \\\cline{3-7}
         & BM 1          & 500  & $4.693$ & $4.978$ & $8.301$ & $21.571$  \\\cline{3-7}
         &               & 1000 & $4.995$ & $5.303$ & $8.928$ & $23.755$  \\\cline{3-7}
         &               & 3000 & $5.467$ & $5.812$ & $9.920$ & $27.306$  \\ \cline{2-7}
         &               & 139  & $4.013$ & $4.246$ & $6.908$ & $16.924$  \\\cline{3-7}
         &               & 300  & $4.351$ & $4.609$ & $7.596$ & $19.197$  \\\cline{3-7}
         & BM 2          & 500  & $4.574$ & $4.850$ & $8.056$ & $20.731$  \\\cline{3-7}
         &               & 1000 & $4.877$ & $5.175$ & $8.683$ & $22.894$  \\\cline{3-7}
         &               & 3000 & $5.350$ & $5.686$ & $9.675$ & $26.421$  \\ \cline{2-7}
         &               & 139  & $4.060$ & $4.297$ & $7.003$ & $17.233$  \\\cline{3-7}
         &               & 300  & $4.398$ & $4.660$ & $7.692$ & $19.502$  \\\cline{3-7}
         & BM 3          & 500  & $4.621$ & $4.901$ & $8.153$ & $21.062$  \\\cline{3-7}
         &               & 1000 & $4.924$ & $5.225$ & $8.780$ & $23.233$  \\\cline{3-7}
         &               & 3000 & $5.396$ & $5.736$ & $9.772$ & $26.770$  \\ \cline{2-7}
         &               & 139  & $4.118$ & $4.359$ & $7.121$ & $17.616$  \\\cline{3-7}
         &               & 300  & $4.456$ & $4.722$ & $7.811$ & $19.902$  \\\cline{3-7}
         &BM 4, 6, and 7 & 500  & $4.679$ & $4.963$ & $8.272$ & $21.472$  \\\cline{3-7}
         &               & 1000 & $4.981$ & $5.288$ & $8.900$ & $23.654$  \\\cline{3-7}
         &               & 3000 & $5.453$ & $5.797$ & $9.891$ & $27.202$  \\ \cline{2-7}
         &               & 139  & $4.094$ & $4.333$ & $7.072$ & $17.457$  \\\cline{3-7}
         &               & 300  & $4.432$ & $4.696$ & $7.761$ & $19.736$  \\\cline{3-7}
         & BM 5          & 500  & $4.655$ & $4.937$ & $8.223$ & $21.302$  \\\cline{3-7}
         &               & 1000 & $4.958$ & $5.262$ & $8.850$ & $23.479$  \\\cline{3-7}
         &               & 3000 & $5.430$ & $5.772$ & $9.842$ & $27.023$  \\ \cline{2-7}
         &               & 139  & $3.950$ & $4.178$ & $6.781$ & $16.520$  \\\cline{3-7}
         &               & 300  & $4.288$ & $4.541$ & $7.467$ & $18.753$  \\\cline{3-7}
         & BM 8          & 500  & $4.511$ & $4.782$ & $7.926$ & $20.294$  \\\cline{3-7}
         &               & 1000 & $4.815$ & $5.108$ & $8.553$ & $22.443$  \\\cline{3-7}
         &               & 3000 & $5.289$ & $5.620$ & $9.546$ & $25.956$  \\ \hline
    \end{tabular}
    \label{HE-HL-bounds}
\end{table*}

\section{\label{dc} conclusions}
In this work, we derived LHC bounds on two different 3-3-1 models, namely 3-3-1 RHN and 3-3-1 LHN. We assessed the impact of overlooked exotic $Z^\prime$ decays in the derivation of lower mass limits using dilepton data. Later, we obtained solid lower mass bounds that range from $3.9$~TeV to $4.1$~TeV, significantly weaker than previous studies. We also forecasted HL-LHC, HE-LHC and FCC-hh mass reach, and put our results into perspective with dark matter phenomenology to conclude that one could successfully accommodate a few TeV thermal dark matter candidate in agreement with direct detection and collider bounds. Our main results are summarized in \tab{HE-HL-bounds}.

\acknowledgments
FSQ thanks Universidad Tecnica Frederico Santa Maria for the hospitality during the final states of this work. We thank Alfonso Zerwekh, Antonio Carcamo, and Carlos Pires for discussions. Y.S.V acknowledges the financial support of CAPES under Grant No. 88882.375870/2019-01. Y.M.O.T. acknowledges financial support from CAPES under grants 88887.485509 / 2020-00. FSQ is supported by ICTP-SAIFR FAPESP grant 2016/01343-7, CNPq grants 303817/2018-6 and 421952/2018 – 0, and the Serrapilheira Foundation (grant number Serra - 1912 – 31613). LCD thanks Simons Foundation (Award Number: 884966, AF) for the financial support. S.K. is supported by ANID PIA/APOYO AFB180002 (Chile) and 
by ANID FONDECYT (Chile) No. 1190845. F.S.Q. and S.K. acknowledge support from  ANID$-$Millennium Program$-$ICN2019\_044 (Chile). AA is supported by CNPq (307317/2021-8). AA and FSQ also acknowledge support from FAPESP (2021/01089-1) grant.

\nocite{*}
\bibliography{ref}%

\end{document}